\documentclass[10pt,conference]{IEEEtran}

\usepackage{booktabs} 
\usepackage{adjustbox}
\usepackage{amsmath}
\usepackage{array, booktabs, makecell}
\usepackage{flexisym}
\usepackage{soul}
\usepackage{balance}
\usepackage{comment}
\usepackage{xcolor}

\ifCLASSINFOpdf
\else
\fi
\begin{document}

\title{ALFAA: Active Learning Fingerprint Based Anti-Aliasing for Correcting Developer Identity Errors in Version Control Data}

\author{

\IEEEauthorblockN{
Sadika Amreen\IEEEauthorrefmark{1},
Audris Mockus\IEEEauthorrefmark{2}}
\IEEEauthorblockA{Electrical Engineering and\\Computer Science\\
University of Tennessee\\
Knoxville, Tennessee\\
\IEEEauthorrefmark{1}samreen@vols.utk.edu,\\
\IEEEauthorrefmark{2}audris@utk.edu
}

\and

\IEEEauthorblockN{
Chris Bogart}
\IEEEauthorblockA{School of Computer Science\\
Carnegie Mellon University\\
Pittsburgh, PA\\
cbogart@cmu.edu}

\and

\IEEEauthorblockN{
Yuxia Zhang}
\IEEEauthorblockA{Electronics Engineering\\ and Computer Science\\
Peking University\\
Beijing, China\\
yuxiaz@pku.edu.cn}

\and

\IEEEauthorblockN{
Russell Zaretzki}
\IEEEauthorblockA{Business Analytics \\ and Statistics\\University of Tennessee\\
Knoxville, Tennessee\\
rzaretzk@utk.edu}

}












\maketitle

\begin{abstract}
  Graphs of developer networks are important for software
  engineering research and practice. For these
  graphs to realistically represent the networks, accurate developer identities are imperative. We aim to identify developer identity errors
  amalgamated from open source software repositories generated by
  version control systems, investigate the nature and prevalence of
  these errors, design corrective algorithms, and estimate the
  impact of the errors on networks inferred from this data.  We
  investigate these questions using a collection of over 1B Git commits
  with over 23M recorded author identities. By inspecting the author
  strings that occur most frequently, we group identity errors into
  categories. We then augment the author strings with three
  behavioral fingerprints: time-zone frequencies, the set of files
  modified, and a vector embedding of the commit messages. We create
  a manually validated set of identities for a subset of OpenStack
  developers using an active learning approach and use it to fit
  supervised learning models to predict the identities for the
  remaining author strings in OpenStack. We then compare these
  predictions with a competing commercially available effort and
  a leading research method. Finally, we compare network measures for
  file-induced author networks based on corrected and raw data.  We
  find commits done from different environments, misspellings,
  organizational ids, default values, and anonymous IDs to be the
  major sources of errors. We also find supervised learning methods
  to reduce errors by several times in comparison to existing research and
  commercial methods and the active learning approach to be an
  effective way to create validated datasets.  Results also
  indicate that correction of developer
  identity has a large impact on the inference of the social
  network.  We believe that our proposed Active Learning Fingerprint Based Anti-Aliasing (ALFAA) approach
  will expedite research progress in the software engineering domain
  for applications that depend upon graphs of developers or other
  social networks.
\end{abstract}

\begin{IEEEkeywords}
Software Repository Mining, Identity Disambiguation,
Random Forest Classification, Record Linkage, Behavioral Fingerprinting, Social Network Analysis
\end{IEEEkeywords}

\newcommand{\keyword}[1]{\textbf{#1}}
\newcommand{\tabhead}[1]{\textbf{#1}}
\newcommand{\code}[1]{\texttt{#1}}
\newcommand{\file}[1]{\texttt{\bfseries#1}}
\newcommand{\option}[1]{\texttt{\itshape#1}}

\section{INTRODUCTION}
\label{intro}

With the advent of the Internet, the study of interactions among
people involved in social networks has become a topic of great
importance and increasing interest.  As software engineering is a
collaborative activity with developers and teams working together on
open source and commercial software projects, much research studies
social networks data in order to provide
insights~\cite{taskBasedSocial} and practical
tools~\cite{czerwonka2013codemine,mockus2002expertise}.  The
analysis of such social interactions can help us understand diverse
concepts such as developer collaboration~\cite{taskBasedSocial}, the
contributions of companies to open source software
projects~\cite{jergensen2011onion,zhou2016inflow}, predicting faults
in software~\cite{Pinzger:2008:DNP:1453101.1453105}, measuring
developer productivity and
expertise~\cite{cataldo2008socio,mockus2009succession}, among
numerous other examples. Furthermore, understanding developer
networks plays a crucial part in the development of software
products and processes~\cite{SoftEngIssues} and improves the
understanding of how open source communities evolve over
time~\cite{Martinez}.

All of this research, however, requires us to determine developer
identities accurately, despite the problematic operational data
extracted from software repositories~\cite{M14}, which contains
incorrect and missing values, such as multiple or erroneous
spellings, identity changes that occur over time, group identities,
and other issues. Identity resolution to identify actual developers
based on data from software repositories is a major challenge mainly
due to (1) the lack of ground truth - absence of validated maps from
the recorded to actual identities and (2) the very large volumes of
data - millions of developer identities in hundreds of millions of code commits.  Incorrect identities are likely to result in
incorrect networks~\cite{wang2012measurement}, thus making the
subsequent analysis and conclusions questionable.

These issues have been recognized in software
engineering~\cite{german2003automating,Bird:2006:MES:1137983.1138016}
and beyond~\cite{Cohen03acomparison}. To cope, studies in the
software engineering field tend to focus on individual projects or
groups of projects where the number of IDs that need to be
disambiguated is small enough for manual validation and devise a
variety of heuristics to solve this formidable problem. The social
networks of scientific paper authors or patents~\cite{nonstars} must
handle a much larger set of identities and the population
census~\cite{Cohen03acomparison} has an even larger set of
identities. The latter literature refers to the accurate identity
problem as record matching: how to match records in one table (e.g.,
the list of actual developers) with records in another table (e.g.,
code commits) where some of the fields used for matching may differ
for the same entity. The way these techniques are applied in
practice for author resolution, however, is primarily to resolve
synonyms (instances where the same person may have multiple ids),
but not homonyms (instances where the same ID is used by multiple
individuals). In software engineering and, in particular, in code
commits to a version control system, author information recorded in
a commit is often reused for multiple individuals: for example, logins (``root'', ``Administrator''), group or tool ids (``Jenkins
Build''), or identifiers sought to preserve anonymity (``John Doe'',
``name@domain.com''). Furthermore, software data does not have
database structure similar to population census where the birth-date
field helps to resolve homonyms, and, more generally, appears to
contain a substantially larger fraction of records with errors.
We address these shortcomings, 
as the following research questions:
\begin{enumerate}
\item What are the most common reasons for identity errors in version control data?
\item What information besides spelling of author names and contacts might help assign authorship to a code commit?
\item How can we minimize the amount of manual effort needed to create accurate identity assignments?
\item How does our approach compare to matching techniques in
  research and commercial efforts applied in a software engineering
  context?
\item What is the impact of identity errors on actual collaboration networks among developers? 
\end{enumerate}

While there has been a lot of progress in disambiguating identities
in authorship for research papers and patents
\cite{Cohen03acomparison,nonstars,Winkler06overviewof,blocking}, the
identity errors in software engineering context appear to be quite
different and we seek a better understanding of their nature and
extent to tailor the correction techniques for the software
engineering domain.

Most traditional record matching techniques use string similarity of
identifiers (typically login credentials) i.e. name, username and
email similarity. A broad spectrum of approaches ranging from direct
string comparisons of name and email~\cite{LinkingAccounts} to
supervised learning based on string similarity~\cite{nonstars} have
been used to solve the identity problem. However, such methods do
not help with homonyms, which are   common in
software engineering data. We, therefore, need additional pieces of
information (an analog to date of birth for census records).  For
this purpose we propose to enhance the string similarity-based
techniques with what we call behavioral fingerprints or activity
patterns that tend to be more similar if different IDs are used by
the same individual and less similar for IDs of distinct individuals. In
our evaluation, we use files modified, the similarity of the text in
commit messages, and the distribution of time zones in code commits
as the three behavioral fingerprints.
The most accurate record matching techniques use supervised machine learning methods. Since we lack a large corpus needed to train such
methods we propose to use active learning
approaches~\cite{Sarawagi:2002:IDU:775047.775087} that allow the
creation of a large validated set of identities with a minimal
amount of manual effort by focusing the manual effort on instances
where the learner has the largest uncertainty.

We compare the accuracy of ALFAA on 16K OpenStack
contributors to a commercially funded effort and to one of the
recent research methods. We also demonstrate that it scales to a
larger dataset of 2M contributors to several large software
ecosystems.
Finally, we assess how identity errors affect file-induced developer
collaboration networks~\cite{wang2012measurement}.
We find that typos, application defaults, organizational
ids, and desire for anonymity are primary cause of errors in
developer identity within a very large body of 1B commits. 
The proposed behavioral
fingerprints improve the accuracy of the
predictor even with a limited training sample. Finally, we find
that the commercial and recent research-based identity
resolution methods for the OpenStack problem have much lower
accuracy than our proposed method and that the errors in the actual
identity data in OpenStack strongly impact the social network
measures. The identity errors represent a real problem that is
likely to affect results of many analysis or development tools, but
these errors can be addressed even for very large datasets using the
proposed approach.

The novelty of our contribution first involves behavioral
fingerprinting that includes Doc2Vec method to find similarities
among commit messages thus providing authorship likelihood measures
even for commits with empty or generic author string.  Second, we
propose the use of machine learning methods in identity resolution
within software engineering context that improve accuracy to a level
comparable or higher than manual matching. This is a radically
different approach from the current state of the art of manually
designed heuristics. The trained models can be further improved
simply by adding larger training sample instead of requiring effort
intensive design and application of customizable heuristics. Models
and data will be shared upon publication.  Third, we propose to use
active learning to minimize effort to generate training
samples. Fourth, we identify several new sources of errors in
developer identity. Fifth, we evaluate accuracy of our approach on a
large sample of 16K OpenStack contributors and compare it to a
commercial method and a recent research method on an extremely large
sample of 2M contributors in large ecosystems.

The remainder of this article is organized as
follows. Section~\ref{relatedWork} discusses the current
state-of-the-art practices in the domain of identity disambiguation. 
Section~\ref{dataSources} discusses the data collection
process and its overview. Section~\ref{classifyingErrors}
discusses the nature of errors associated with
developer identities as well as their reasons.  
Section~\ref{approach} discusses the
approach in solving the identity disambiguation problem by
correcting synonym and homonym errors and results from ALFAA. Section~\ref{evaluation} compares the
results produced by ALFAA to a commercial effort and
recent research method. Section~\ref{networkError} demonstrates the
impact of identity errors on networks by using a developer
collaboration network and finally, Section~\ref{conclusion}
summarizes findings and provides conclusions.

\vspace{-2mm}
\section{RELATED WORK}
\label{relatedWork}

The issue of identity resolution through disambiguation or de-anonymization falls under the broader field of "Record Linkage". The first mathematical model for record linkage introduced in 1969 by Ivan Fellegi and Alan Sunter~\cite{FellegiSunter}, is used to identify duplicates when unique identifiers are unavailable. This model serves as the basis of many record linkage methods practiced today. Since then, the problem has been investigated in many fields such as on patent data~\cite{nonstars} to link records of the companies, organizations and individuals or government agencies to which a patent is assigned, on US census data~\cite{Winkler06overviewof}, synthetic census data~\cite{Cohen03acomparison} and in the construction of web services that integrate crowd-sourced data such as CiteSeer~\cite{citeseer}. It has also been used in the field of empirical software engineering research~\cite{german2003automating,Bird:2006:MES:1137983.1138016} to disambiguate identities of people in a software ecosystem for various purposes such as to build social diversity dataset from thousands of GitHub projects~\cite{msrData15}, to assess the contributor’s total activity
within projects~\cite{Gharehyazie2015} in Open Source Software and across platforms~\cite{GH_SO} and in mailing lists~\cite{MailingList}. Most of these are still reliant on simple string matching heuristics.
The issue of developer identities has been a serious problem in software repository mining, particularly when trying to combine information from different types of data sources in a coherent way where the available data concerning persons involved in a project may be dispersed across different repositories~\cite{GOEMINNE2013971,Robles:2005:DIM:1083142.1083162}.

Approaches such as merging identities with similar name labels, email addresses or any combination of these have been used in the past for disambiguation. For example, an algorithm~\cite{Bird:2006:MES:1137983.1138016} designed specifically to detect identities belonging to developers who commit to code repositories and people participating in a mailing list uses string similarity based on Levenshtein distance on first, last, and user name fields of developers and mailers coupled with a threshold parameter. This assumes a name will be split into two parts using whitespace or commas as delimiters and user names can be derived from the email address string. This algorithm was later modified to include more characters as separators, extended to account for an arbitrary number of name parts and include more individuals from bug repositories and then evaluated using different identity merge algorithms~\cite{GOEMINNE2013971}.
While these approaches are reported to perform well only through string matching and thresholding, for example, work using more sophisticated heuristics such as Latent Semantic Analysis (LSA) on names of GNOME Git authors which was also used for disambiguation~\cite{Gnome}, fail to address issues where developer identity strings are problematic, i.e. incomplete or missing.

Other research on the data from the U.S. patent and trademark (USPTO)~\cite{nonstars} database uses a supervised learning approach based on a large set (over 150,000) of hand labeled inventor records to perform disambiguation. This, therefore, assumes an availability of sufficient and reliable ground truth data to perform a supervised learning approach. A major challenges we face with disambiguation is the lack of an adequate pool of hand labeled data to use for supervised learning. Furthermore, these prior approaches fail to address the problem of homonyms resolution i.e. where a single label may be used by multiple identities. This is critical because excluding problematic nodes from a network can radically alter the properties of the social network as well as nodes (e.g., developer productivity, tenure with the project, etc).

The fact that there is insufficient ground truth for our dataset of developers from projects hosted on GitHub causes a hindrance to employing any supervised learning approach directly. Past research on de-duplication of authors in citations~\cite{Sarawagi:2002:IDU:775047.775087} has leveraged a technique called active learning, which starts with limited labels and a large unlabeled pool of instances, thereby, significantly reducing the effort in providing training data manually. The active learning method uses an initial classifier to predict on some unlabeled instances. The initial classifier produces some results (a higher fraction) with high confidence and some others (a lower fraction) with lower confidence i.e. the classifier's confusion region. This confusion region can therefore be extracted and manually labeled for it to serve as the training data for the actual classifier.  


In summary, the current state of art in software engineering remains
based on designing a set of matching heuristics with manual verification and techniques from other fields need tailoring for the types of problems common in software engineering. We propose an approach that addresses these shortcomings and that could be combined with other, more specialized approaches, especially for resolving homonyms more precisely. For example, the productivity outlier detection and reallocation approach~\cite{Qimu} detects when the number of commits or changes is highly unusual and distributes the authorship to other committers. Such an approach would help to both identify homonyms and redistribute authorship to each developer. 

\vspace{-3mm}
\section{DATA SOURCES}
\label{dataSources}

Version Control System (VCS) is an ubiquitous tool in software
development and it tracks code modifications (commits). Each time a
new commit is made, the VCS records authorship, commit time, commit
message, parent commit and the full folder structure after the
commit. Author string in a commit consists of the author name (first
and last) and their email addresses. We determine files modified in
a commit by comparing the full folder structure prior to and after
the commit.  We have been collecting such data from projects with
public VCS since 2007~\cite{msr09} and currently have 1.1B commits
made by over 20M authors in 46M VCS repositories.

We set out to find a subset of this data that includes a sizable
set of projects where we could compare the results not only to
research-based methods but also to approaches used in industry. We,
therefore, selected the OpenStack ecosystem as it already had an
implementation of disambiguation by Bitergia, a commercial firm,
which mapped multiple developer IDs to an unique identifier
representing a single developer as well as mapping contributors to
their affiliated companies.

OpenStack\footnote{https://www.openstack.org/} is a set of software
tools for building and managing cloud computing platforms for both
public and private clouds. It lets users deploy virtual machines and
can handle different tasks for managing a cloud environment on the
fly\footnote{https://opensource.com/resources/what-is-openstack}. We
discovered 1,294 projects that are currently hosted on GitHub and
have 16,007 distinct author strings in the associated
commits. Moreover, to measure the scalability of our method, we selected an even larger collection of projects from several large open source ecosystems having approximately 2M developer identities.

{\small
\begin{table*}[htb]
\centering
\caption{Data Overview: The 10 most frequent names and emails}
\label{tab:Freq}
\begin{adjustbox}{width=0.9\textwidth}
\begin{tabular}{l l l l l l l l l l}
\toprule
\tabhead{Name} & \tabhead{Count} & \tabhead{First Name} & \tabhead{Count} & \tabhead{Last Name} & \tabhead{Count} & \tabhead{Email} & \tabhead{Count} & \tabhead{User Name} & \tabhead{Count} \\
\midrule
unknown & 140859 & unknown & 140875 & unknown & 140865 & \textless{blank}\textgreater & 16752 & root & 72655\\
root & 66905 & root & 66995 & root & 67004 & none@none & 9576 & nobody & 35574\\
nobody & 35141 & David & 45091 & nobody & 35141 & devnull@localhost & 8108 & github & 19778\\
Ubuntu & 18431 & Michael & 40199 & Ubuntu & 18560 & student@epicodus.com & 5914 & ubuntu & 18683\\
(no author) & 6934 & nobody & 35142 & Lee & 10826 & unknown & 3518 & info & 18634\\
nodemcu-custom-build & 6073 & Daniel & 34889 & Wang & 10641 & you@example.com & 2596 & \textless{blank}\textgreater & 17826 \\
Alex & 5602 & Chris & 29167 & Chen & 9792 & anybody@emacswiki.org & 2518 & me & 14312\\
System Administrator & 4216 & Alex & 28410 & Smith & 9722 & = & 1371 & admin & 12612\\
Administrator & 4198 & Andrew & 26016 & Administrator & 8668 &  Unknown & 1245 & mail & 11253\\
\textless{blank}\textgreater & 4185 & John & 25882 & User & 8622 &  noreply  & 913 & none & 11004\\
\bottomrule\\
\end{tabular}
\end{adjustbox}
\vspace{-6mm}
\end{table*}
}

\section{CLASSIFYING ERRORS}
\label{classifyingErrors}

In order to tailor existing identity resolution approaches (or create new ones), we need a better understanding of the nature of the errors
associated with the records related to developer identity. For example, in census data a common error may be a typo, a variation in the phonetic spelling of a name, or the reversal of the first and last names, among others. Previous studies have identified errors as a result of transliteration, punctuation, irrelevant information incorporated in names, etc.~\cite{Gnome,PersonalName}.
Furthermore, complications 
are at times introduced by the use of tools. Author information in a Git commit (which we study here) depends on an entry specifying user name and email in a Git configuration file of the specific computer 
a developer is using at the moment. Once Git commit is recorded, it is immutable like other Git objects, and cannot be changed. Once a developer pushes their commits from the local to remote repository, that author information remains. A developer may have multiple laptops, workstations, and work on various servers, and it is possible and, in fact, likely, that on at least one of these computers the Git configuration file has a different spelling of their name and email. 
It is not uncommon to see the commits done under an organizational alias, thus obscuring the identity of the author. 

Some Git clients may provide a default value for a developer, for example, the host name. Sometimes developers do not want their identities or their email address to be seen, resulting in 
intentionally anonymous name, such as, John Doe or email, such as devnull@localhost. Developers may change their name over time, for example, after marriage, creating a synonym and other scenarios may be possible. 

In order to correct this, we need to determine the common reasons
causing errors to be injected into the system. We therefore,
inspected  authors strings from our collection of over 1B commits. 
First, we inspected random subsets of author IDs to understand how or
why these errors occur. We then inspected the most common names and
user names  and determined that many of them were unlikely to be names of individuals. We also came across many additional types of errors when we manually labeled our data in the active learning phase as we discuss in Section~\ref{approach}. We identify these errors and broadly categorize them into synonym and homonym  errors.

\textbf{Synonyms:} These kinds of errors are introduced when a person
  uses different strings for names, user-names or email
  addresses. For example, `utsav dusad 
 $<$utsavdusad @gmail.com$>$' and `utsavdusad   $<$utsavdusad@gmail.com$>$' are identified as
  synonyms.  
Spelling mistakes such as `Paul Luse
$<$paul.e.luse @intel.com$>$' and `paul luse
$<$paul.e.luse@itnel.com$>$' are also classified as
synonyms, as `itnel' is likely to be a misspelling of `intel'.

\textbf{Homonyms:} Homonym errors are introduced when multiple people
  use the same organizational email address. For example, the ID
  `saper $<$saper@saper.info$>$' may be used by
  multiple entities in the organization. For example 'Marcin Cieslak
  $<$saper@saper.info$>$' is an entity who may have
  committed under the above organizational alias.  
Template credentials from tools is another source that might
introduce homonym errors in the data as some users may not enter
values for name and/or an email field such as `Your Name
$<$vponomaryov@mirantis.com$>$' may belong to
the following author - `vponomaryov
$<$vponomaryov@mirantis.com$>$'. This may be due to
the user's desire for anonymity. Generic names such as John Doe,
 me@email.com, a@b.com add to homonym error as well. Homonym errors
 are also introduced when a user leaves the name or email field
 empty, for example, `chrisw \textless{unknown}\textgreater'. A brief frequency analysis showed that the most frequent names in the dataset
such as 'nobody', `root', and 'Administrator' are a result of
homonym errors as shown in Table~\ref{tab:Freq}\footnote{We provide
  the actual email and name of individuals as found in the commits
  for the submitted version of the manuscript but will change the
  names and email to randomly selected elements from a large dataset
  of first names, last names, and username(s) for the published
  version in order to preserve privacy of these individuals.}. 


\vspace{-1mm}
\section{DISAMBIGUATION APPROACH}
\label{approach}

Following traditional record linkage methodology and identity
linking in software~\cite{Bird:2006:MES:1137983.1138016} we first
split the information in the author string into  
several fields and define similarity metrics for all author pairs. 
We also incorporate the term frequency measure for each of the attributes in a pair. Finally, we add similarity 
between behavioral fingerprints. We generate a table of these similarity measures for all 256,224,049 author pairs in the 
OpenStack dataset. 

\vspace{-1mm}
\subsection{String Similarity Measures}
\label{strSim}

Each author string is stored in the following format - ``name \textless{email}\textgreater'', e.g. ``Hong Hui Xiao \textless{xiaohhui@cn.ibm.com}\textgreater''. We define the following attributes for each user.

\begin{enumerate}
\item Author: String as extracted from source as shown in the example above
\item Name: String up to the space before the first `\textless'
\item Email: String within the `\textless\textgreater' brackets
\item First name: String up to the first space, `+', `-', `\_', `,', `.' and camel case encountered in the name field
\item Last name: String after the last space, `+', `-', `\_', `,', `.' and camel case encountered in the name field 
\item User name: String up to the `@' character in the email field
\end{enumerate}

Additionally, we introduce a field `inverse first name' whereby in
the comparison between two authors it is compared to the last name
in the other record.  We introduce this field to make sure that our
algorithm captures cases where authors reverse the order of their
first and last names.  In the case where there is a string without
any delimiting character in the name field, the first name and last
name are replicated. For example, bharaththiruveedula
\textless{bharath\_ves@hotmail.com}\textgreater would have
`bharaththiruveedula' replicated in the first, last and the name
field.  We calculate both Levenshtein and the Jaro-Winkler
similarity as we have seen in previous studies~\cite{Bird:2006:MES:1137983.1138016,Gnome}, which are standard measures for string similarity, for each author pair. To do this, we use an existing implementation of the measures in the RecordLinkage~\cite{recordLinkage} package in R, namely the levenshteinSim() and jarowinkler()  functions. In a
preliminary investigation, we found that the Jaro-Winkler similarity
produces better scores which are more reflective of
similarity between author strings than the Levenshtein score and,
therefore, use this measure in the proposed method.
The Jaro Similarity is defined as 
\[
    sim_j= 
\begin{cases}
	0, & \text{if } m = 0 \\
    \dfrac{1}{3} \left(\dfrac{m}{\vert{s_1}\vert} + \dfrac{m}{\vert{s_2}\vert} + \dfrac{m-t}{m}\right)  & \text{otherwise} \\
\end{cases}
\]
where \textit{s\textsubscript{i}} is the length of string \textit{i}, \textit{m} is the number of matching characters and \textit{t} is half the number of transpositions.
The Jaro-Winkler Similarity modified the Jaro similarity so that differences at the beginning of the string has more significance than differences at the end. 

\subsection{Adjustment Factors for String Frequency}
If two author IDs
share an uncommon name that gives greater confidence than the IDs that share a common name such as
``John''. Furthermore, certain names like ``nobody'' or ``root'' do
not carry any information about the authorship and should be disregarded
in the similarity detection. This
extra information, if properly encoded, could be exploited by a machine learning
algorithm making disambiguation decisions. 
We, therefore, count the number of occurrences of the attributes for each author
as defined in Section~\ref{strSim} i.e. name, first name, last name,
user name and email for our dataset. We calculate the similarity
between author pairs, authors a\textsubscript{1} and
a\textsubscript{2}, for each of these attributes as follows: 
\vspace{-1mm}
\[
    f_{sim}= 
\begin{cases}
	\log _{10}\dfrac{1}{f_{a_1} \times f_{a_2}} & \text{if } a_1~\text{and } a_2 \text{ are valid} \\
    -10 & \text{otherwise }  \\
\end{cases}
\]
where f\textsubscript{a1} and f\textsubscript{a2} are the frequency of names of authors a\textsubscript{1} and a\textsubscript{2} respectively.
We generate a list of 200 common strings of names, first names, last
names and user names and emails from the full dataset of
authors (the first 10 shown in Table~\ref{tab:Freq}) and manually
remove names that appear to be non-fictitious, i.e. names that could truly belong to a person such as Lee, Chen, Chris,
Daniel etc. We set string frequency similarity of a pair of name or
first name or last name or user name to -10 if at least one element
of the pair belongs a string identified as fictitious. This was
done in order to let the learning algorithm recognize the difference
between the highly frequent strings and strings that are not useful
as author identifiers. -10 was chosen because we found that the value for other highly frequent terms were significantly greater. 

\vspace{-2mm}
\subsection{Behavioral Fingerprints}
\label{fingerprint}

In addition to the spelling of the name and contact information,
developers might leave their signature in the way they compose
commit messages, the files they commonly modify, or the time zones
they work in. We designed three similarity measures 
to encode the behavioral attributes of authors - (1)
Similarity based on files touched --- two author IDs
modifying similar sets of files are more likely to represent
the same person. (2) Similarity based on
time zone --- two author IDs committing in the same time zone
indicate geographic proximity and, therefore, higher likelihood
of being the same individual. (3) Similarity based on commit message
text --- two author IDs sharing writing style and vocabulary
increase chances that they represent the same entity.
Operationalizations of these behavioral fingerprints
are given below. 

\textbf{Files modified}: Each modified file is inversely
  weighted using the number of  distinct authors who have modified it (for
  the similar reasons common names are down-weighted as evidence of
  identity). The pairwise
  similarity between authors, $a_1$ and
  $a_2$, is derived by adding the weights of the
  files, W\textsubscript{f}, touched by both authors. A similar metric was found to work
  well for finding instances of succession (when one developer takes
  over the work of another developer)~\cite{M09}. The weight of a file is defined as follows where A\textsubscript{f} is a set of authors who has modified file \textit{f}.
\vspace{-1mm}
\begin{displaymath}
	W_f=\dfrac{1}{A_f}, \text{where } A_f=\vert{a^f_1,..., a^f_n}\vert
\end{displaymath}
\begin{displaymath}
  Sim_{a_1a_2} = \sum_{i=1}^{n_{a_1a_2}}W_{f_i}, \text{where } n_{a_1a_2} = \vert{f_{a_1}}\cap {f_{a_2}}\vert
\end{displaymath}

\textbf{Time zone}: We discovered 300 distinct time zone
  strings (due to misspellings) from the commits and created a `author by time zone'
  matrix that had the count of commits by an author at a given
  time-zone. All time zones that had less than 2 entries (authors) were
  eliminated from further study. Each author was therefore assigned a
  normalized time-zone vector (with 139 distinct time zones) that represents
  the pattern of his/her commits. Similar to the previous metric, we
  weighted each time zone by the inverse number of authors who
  committed at least once in that time-zone. We multiply each
  author's time zone vector by the weight of the time zone. We
  define author a\textsubscript{i}'s time-zone vector as:  

\vspace{-1mm}
\begin{displaymath}
  (TZV_{a_i}^t) =  \left(\dfrac{C_{a_i}^t}{A_{t}}\right),
\end{displaymath}


Here, $(C^t_{a_i})$ is the vector representing the commits of an
author $a_i$ in the different time zones $t$ and $(A_t)$ is the vector representing the number of authors in the different time zones.
The pairwise similarity metric between author a\textsubscript{1} and author a\textsubscript{2} is calculated using the cosine similarity as:

\begin{displaymath}
  tzd_{a_1a_2} = cos\_sim(TZV_{a_1},TZV{a_2})
\end{displaymath}

where TZV\textsubscript{a1} and TZV\textsubscript{a2} are the authors' respective vectors.

\textbf{Text similarity}: 
We use the Gensim's implementation~\footnote{https://radimrehurek.com/gensim/index.html} of the Doc2Vec~\cite{doc2vec} algorithm to generate
vectors that embed the semantics and style of the commits messages of each author. All commit messages for each individual who contributed at least once to one of the OpenStack projects were gathered from the collection described above and 
a Doc2Vec model was built. We obtained a 200 dimensional vector for each of the 16,007 authors in our dataset and calculated cosine similarity to find pairwise similarity between authors. 

However, there are several potential drawbacks of these distance
metrics. For example, high scores for files touched may mean that
two different individuals are working on the same project thereby
editing the same files at alternating times. High document
similarity may mean that the authors share similar vocabulary in the
commit messages which may also be influenced by the work on the same
project. Fortunately our approach leaves the decision to include a
specific feature to a machine learning algorithm. As
we show later in the results, the behavioral similarity measures,
in particular, text similarity, are important predictors for
disambiguation. 

\vspace{-2mm}
\subsection{Data Correction}

\begin{figure*}
\centering
\includegraphics[width = 0.8\textwidth]{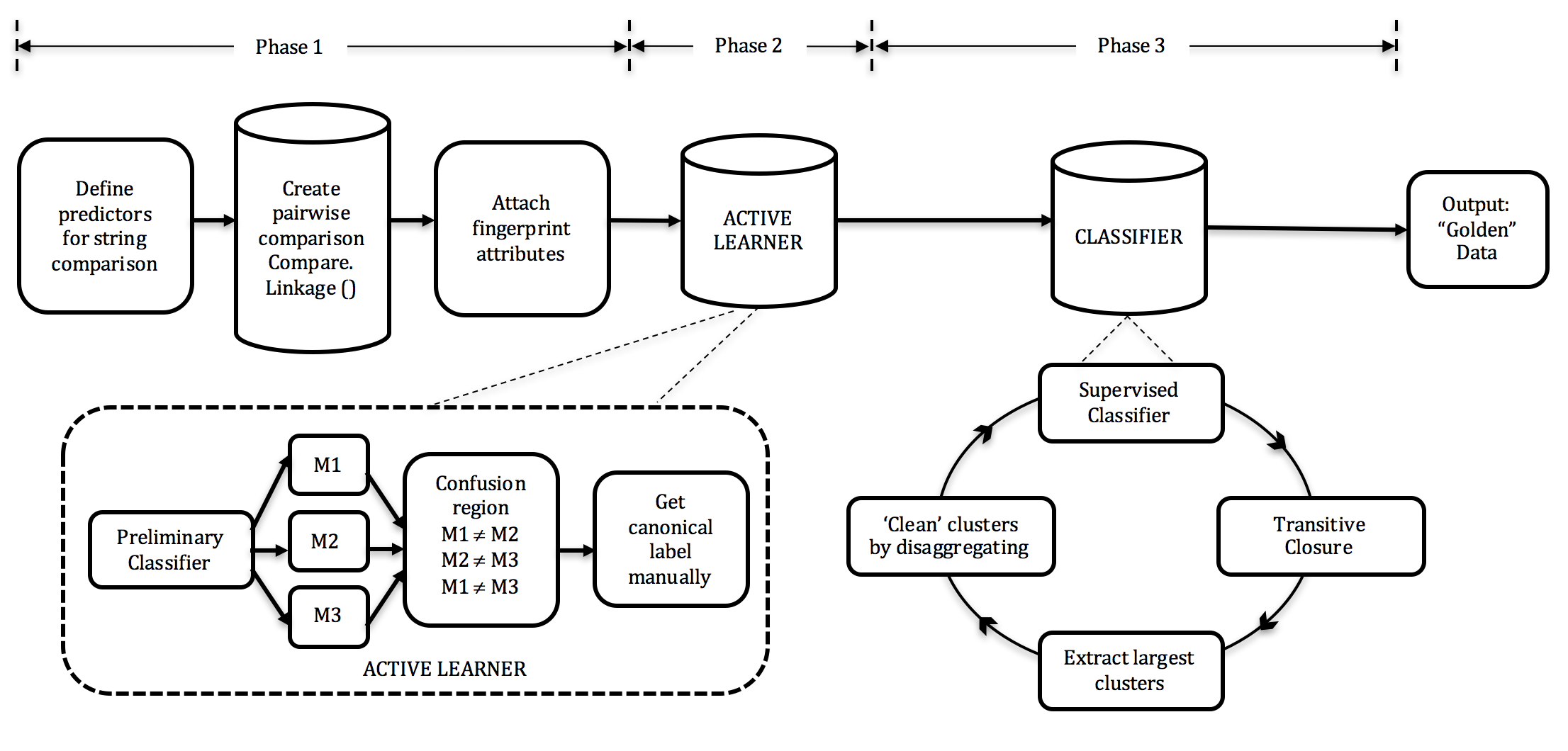}
\caption{Concept of the Disambiguation Process} 
\label{workflow}
\vspace{-6mm}
\end{figure*}

The data correction process can be divided into 3 phases as shown in Figure~\ref{workflow}.
\begin{enumerate}
\item Define predictors - Compute string similarity, frequency similarity and behavioral similarity
\item Active learning - Use a preliminary classifier to extract a
  small set from the large collection of data and generate labels for further classification.
\item Classification - Perform supervised classification, transitive closure, extract clusters to correct, and dis-aggregate incorrectly clustered IDs.
\end{enumerate}

\textbf {PHASE 1: Define Predictors for the Learner} 
Once we have defined the attributes (name, first name, last name,
email, username) for which we want to calculate string similarity,
we use relevant functions implemented in the RecordLinkage
library~\cite{recordLinkage} to obtain the Jaro-Winkler similarity
between each pair of attributes: authors' name, first name, last
name, user name, email and the first author's first name to the 2nd
author's last name (we refer to this as the inverse first name).
In addition to the
string similarities based on these fields, we also include the term
frequency metric, as is commonly done in record matching
literature. The high frequency values tend to carry less
discriminative power than infrequent email addresses or names.
Finally, we include three fingerprint metrics --- files touched,
time-zone, and commit log text. The resulting data is used
as an input to the next phase, i.e. the active learning process.

{\small
\begin{table}
\caption{Confusion region from the preliminary classifier}
\label{tab:predictor}
\centering
\begin{tabular}{l l l}
\toprule
\tabhead{Model1} & \tabhead{Model2} & \tabhead{Model3}  \\
\midrule
Link & Link & No-Link\\
Link & No-Link & Link\\
No-Link & Link & Link\\
No-Link & No-Link & Link\\
No-Link & Link & No-Link\\
Link & No-Link & No-Link\\
\bottomrule\\
\end{tabular}
\vspace{-8mm}
\end{table}
}

\textbf {PHASE 2: Active Learning}
Supervised classification requires ground truth data. As noted
earlier, it is extremely time consuming and error-prone to produce a
large set of manually classified data to serve as an input for a
supervised classifier. Moreover, identifying a small subset of
instances so that the classifier would produce accurate results on
the remainder of the data is also challenging. We use a concept
called \textit{Active Learning}~\cite{Sarawagi:2002:IDU:775047.775087} using a
preliminary classifier that helps us extract a small set of author
pairs that is viable for manual labeling, from the set of over 256M
author pairs.

In this phase, we design the preliminary classifier.
We do a ten-fold
cross validation by first partitioning
the data into ten parts and fit three learners on 
nine parts and predict on the remaining part (prediction set). Each classifier learns from manually classified
pairs and outputs links or non-links for each author pair in the
prediction set. The three classifiers trained on different training
subsets yield  
slightly different predictions (links and no-links for each
pair). The mismatch between predictions of two such classifiers  
indicates instances where the classifier has large uncertainty
(confusion regions) and are shown in Table~\ref{tab:predictor}.  
We conducted manual classification on the cases in
the confusion region of the three classifiers that involved 2,345 pairs.
Each pair was manually inspected and a canonical label (with proper
spelling of the name and email address) was 
selected from among the existing author identities as shown in
Table~\ref{tab:TrainData}.
This step produced a preliminary set of training data with over 2K
pairs for the supervised classification.

{
\begin{table}
\centering
\caption{Example of Training Data}
\label{tab:TrainData}
\begin{adjustbox}{width=\linewidth}
\begin{tabular}{l l}
\toprule
\tabhead{Canonical Label} & \tabhead{Author Identity} \\
\midrule
Jason Koelker \textless{jason@koelker.net}\textgreater & Jason Koelker \textless{jason@koelker.net}\textgreater\\
Jason Koelker \textless{jason@koelker.net}\textgreater & Jason Kölker \textless{jason@koelker.net}\textgreater \\
Tatyana Leontovich \textless{tleontovich@mirantis.com}\textgreater & Tatyana Leontovich \textless{tleontov@yahoo-inc.com}\textgreater \\
Tatyana Leontovich \textless{tleontovich@mirantis.com}\textgreater & Tatyana Leontovich \textless{tleontovich@mirantis.com}\textgreater \\
Tatyana Leontovich \textless{tleontovich@mirantis.com}\textgreater & Tatyana Leontovich \textless{tleontovich@griddynamics.com}\textgreater \\
Tatyana Leontovich \textless{tleontovich@mirantis.com}\textgreater & Tatyanka \textless{tleontovich@mirantis.com}\textgreater\\
Tatyana Leontovich \textless{tleontovich@mirantis.com}\textgreater & TatyanaLeontovich \textless{tleontovich@mirantis.com}\textgreater\\
\bottomrule\\
\end{tabular}
\end{adjustbox}
\vspace{-8mm}
\end{table}
}

A Random Forest model
was then fit using manually classified training data using all 16
attributes (name, email, first name, last name, user name, inverse
first name, name frequency, email frequency, last name frequency,
first name frequency, user name frequency, files touched, time-zone,
and text similarity). We then extracted the importance of each variable in the
model and dropped the attributes with low importance. Upon examining
the incorrect model predictions during the validation stage, we
observed that many of the classifier mistakes were not mistakes
after all. Instead, the classifier was able to identify a number of
mistakes in the manual labeling. We made appropriate correction in
the manually classified data before doing ten-fold cross validation
described in the next section.

\textbf {PHASE 3: Classification}
\label{classification}
Once the labeled dataset is created, we use it to train random
forest models which are commonly used in record matching
literature. We perform a 10-fold cross validation using this
method. The results are shown in Table~\ref{10fold}.  The final
predictor of identity matches involves a transitive closure on the
pairwise links obtained from the classifier\footnote{We found that
  more accurate predictors can be obtained by training the learner
  only on the matched pairs, since the transitive closure typically
  results in some pairs that are extremely dissimilar, leading the
  learner to learn from them and predict many more false positives}.
The result of the transitive closure is a set of connected
components with each cluster representing a single author. Once the
clusters are obtained, we consider all clusters containing 10 or
more elements since a significant portion of such clusters had
multiple developers grouped into a single component. The resulting
20 clusters - 44 elements in the largest and 10 elements in the
smallest cluster among these, were then manually inspected and
grouped.  This manual effort included the assessment of name, user
name and email similarity, projects they worked on, as well as
looking up individual's profiles online where names/emails were not
sufficient to assign them to a cluster with adequate confidence.  An
example of cluster reassignment is given in Table~\ref{tab:disagg}
where we dis-aggregated a single large cluster of 11 IDs to 3
smaller clusters. The first column is the author ID, the second is
the cluster number the ID was assigned to by the algorithm, the
third column is the manually assigned cluster number after
disaggregation. We noticed that, the largest cluster of size of 44
included all IDs that were associated with 'root' and therefore were
not representative of any actual developer. Therefore, we
dis-aggregated the entire cluster to form 44 single element
clusters. The output of this phase is a cleaned dataset in which we
have corrected synonym errors via machine learning and fixed some of
the homonym errors by inspecting the largest clusters. Since an
experiment selecting a sample of pairs from this resulting set and
validating them had very low level error rates, we use it as a
reference or `golden' dataset representing developer identities for
the further analysis.

\vspace{-1mm}
\subsection{Results}
We evaluate the models using the standard measure of correctness -
precision and recall - using the true positive (\textit{tp}), true
negative (\textit{tn}), false positive (\textit{fp}) and false
negative (\textit{fn}) outcomes produced by the models. We obtained
an average precision of 99.9\% and an average recall of 99.7\% from
the  10-fold cross validation of the random forest model shown in
Section~\ref{10fold}. 

\begin{table}
\centering
\caption{Confusion Matrix of 10-fold cross validation of the Random Forest Model}
\label{10fold}
\begin{adjustbox}{width=\linewidth}
\begin{tabular}{l|ll|ll|ll|ll|ll}
\toprule
 & \tabhead{0}    & \tabhead{1}     
 & \tabhead{0}    & \tabhead{1}     
 & \tabhead{0}    & \tabhead{1}     
 & \tabhead{0}    & \tabhead{1}     
 & \tabhead{0}    & \tabhead{1}     \\ 
 \hline

0 & 
549,609 & 4   & 
548,179 & 3   & 
549,469 & 2   & 
551,136 & 5   & 
550,108 & 5  \\

1 & 
0 & 992       & 
2 & 1,110     & 
0 & 1,082     & 
0 & 1,039     & 
3 & 1,014     \\

\toprule
 &  \tabhead{0}    & \tabhead{1}     
 & \tabhead{0}    & \tabhead{1}     
 & \tabhead{0}    & \tabhead{1}     
 & \tabhead{0}    & \tabhead{1}     
 & \tabhead{0}    & \tabhead{1}     \\
 
 \hline
 
0 & 
549,204 & 1   & 
549,402 & 1   & 
547,958 & 4   & 
548,730 & 4   & 
549,569 & 2  \\

1 & 
2 & 1,075     & 
1 & 1,033     & 
0 & 1,021     & 
0 & 1,084     & 
0 & 1,010     \\

\bottomrule
\end{tabular}
\end{adjustbox}
\vspace{-3mm}
\end{table}

\vspace{-1mm}

{
\begin{table}
\centering
\caption{Cluster Cleanup through Manual Disaggregation}
\label{tab:disagg}
\begin{adjustbox}{width=\linewidth}
\begin{tabular}{l l l}
\toprule
\tabhead{Author Identity} & \tabhead{Cluster\#} & \tabhead{New Cluster\#} \\
\midrule
AD \textless{adidenko@mirantis.com}\textgreater & 22 & 1 \\
Aleksandr Didenko \textless{adidenko@mirantis.com}\textgreater & 22 & 1\\
Alexander Didenko \textless{adidenko@mirantis.com}\textgreater & 22 & 1\\
Sergey Vasilenko \textless{stalker@makeworld.ru}\textgreater & 22 & 2 \\
Sergey Vasilenko \textless{sv854h@att.com}\textgreater & 22 & 2\\
Sergey Vasilenko \textless{sv@makeworld.ru}\textgreater & 22 & 2\\
Sergey Vasilenko \textless{svasilenko@mirantis.com}\textgreater & 22 & 2 \\
Sergey Vasilenko \textless{xenolog@users.noreply.github.com}\textgreater & 22 & 2 \\
Vasyl Saienko \textless{vsaienko@mirantis.com}\textgreater & 22 & 3 \\
vsaienko \textless{vsaienko@cz5578.bud.mirantis.net}\textgreater & 22 & 3 \\
vsaienko \textless{vsaienko@mirantis.com}\textgreater & 22 & 3 \\
\bottomrule\\
\end{tabular}
\end{adjustbox}
\vspace{-6mm}
\end{table}
}

Since record matching is a slightly different problem from
traditional classification, the literature introduces two  
additional error metrics: splitting and
lumping~\cite{SplitLump}. Lumping occurs when multiple author IDs
are identified to belong to a single developer.  
The number of lumped records is defined as the number of records
that the disambiguation algorithm incorrectly mapped to the largest
pool of IDs belonging to a given author.  
Splitting occurs when an ID belonging to a single developer is
incorrectly split into IDs representing several physical
entities. The number of split records is defined as the number of
author IDs that the disambiguation algorithm fails to map to the
largest pool of IDs belonging to a given author.
\vspace{-2mm}
{\small
\begin{table}[htbp]
\caption{Largest cluster Corresponding to Single Entity with Highest Aliases After Disaggregation}
\label{tab:largestCluster}
\begin{adjustbox}{width=\linewidth}
\centering
\begin{tabular}{l l}
\toprule
\tabhead{AuthorID} & \tabhead{AuthorID}  \\
\midrule
Greg Holt \textless{gholt@rackspace.com}\textgreater & tlohg \textless{z-github@brim.net}\textgreater  \\
Greg Holt \textless{greg@brim.net}\textgreater & tlohg \textless{gholt@rackspace.com}\textgreater\\
Greg Holt \textless{gregory.holt@gmail.com}\textgreater & gholt \textless{z-launchpad@brim.net}\textgreater\\
Greg Holt \textless{gregory_holt@icloud.com}\textgreater & gholt \textless{z-github@brim.net}\textgreater \\
Greg Holt \textless{z-github@brim.net}\textgreater & gholt \textless{gregory.holt+launchpad.net@gmail.com}\textgreater\\
Gregory Holt \textless{gholt@racklabs.com}\textgreater & gholt \textless{gholt@rackspace.com}\textgreater\\
gholt \textless{devnull@brim.net}\textgreater & gholt \textless{gholt@brim.net}\textgreater\\
\bottomrule\\
\end{tabular}
\end{adjustbox}
\vspace{-6mm}
\end{table}
}

Since, these two metrics only focus on the largest pool of IDs
belonging to a single developer and ignores the other clusters of IDs
corresponding to the same unique developer, the work
in~\cite{nonstars} modifies these measures to evaluate all pairwise
comparison of author records made by the disambiguation
algorithm. According to the latter approach, we create a confusion
matrix of the pairwise links from the golden data 10,950set and the links
created by the classifier.

and calculate splitting and lumping in the
following manner: 
\begin{displaymath}
  Splitting = \dfrac{fn}{tp + fn}, 
  Lumping = \dfrac{fp}{tp + fn}
\end{displaymath}

From the cross-validation, 0.3\% of the cases were split and 0.1\% of the cases were lumped.
We use one of these models to predict links or non-links for our
entire dataset of over 256M pairs of records. The classifier found 31,044
links and we generated an additional 3,293 links through transitive
closure. Therefore, we have 34,337 pairs linked after running the
disambiguation algorithm. Using this, we constructed a network that
had 10,835 clusters that were later manually inspected and
disaggregated using the procedure described in
subsection~\ref{classification}. Finally, we were left with 10,950
clusters, each representing an author, with 14 elements in the largest cluster, corresponding to the highest number of aliases by a single individual as shown in Table~\ref{tab:largestCluster}.


\vspace{-1mm}
\section{EVALUATION}
\label{evaluation}

In this section we try to answer questions related to the accuracy
of the manually labeled training data and to compare our approach to
two alternatives from the commercial and research domains. It is
important to note that we are evaluating our algorithm trained on a
small amount of training data and, as with other machine learning
techniques, we expect it to have higher accuracy with more training
data that would be added in the future.
\vspace{-2mm}
\subsection{Accuracy of the training data}

The absence of ground truth requires us to investigate the accuracy
of the training data.  Two independent human raters (authors who are PhD students in
Computer Science) were presented with the spreadsheet containing 
1060 pairs of
OpenStack author IDs and marked it using the following protocol.  Each rater was
instructed to inspect each pair of author IDs (full name and email)
listed in the spreadsheet and supplemented by author's affiliations (see
Section~\ref{bitergia}, the dates of their first and last commits in
the OpenStack projects, and their behavioral similarity scores.
Each rater was instructed to mark author pair as a match (1) if the
two identities are almost certainly from the same person, a
non-match (0) if they are certainly not from the same person, and
provide a number in between zero and 1 reflecting the raters
subjective probability that they are representing the same person.
Each rater was instructed to use the above mentioned information
(listed next to the pair in the spreadsheet) to make their decision
and were instructed to search for developers on github or google if
they did not feel confident about their decision.
For cases where both raters marked either zero or one we found 1011
instances of agreement and 17 cases of disagreement between the two
raters. By thresholding the 32 cases that had probability value
greater than zero and less than one to the nearest whole, we obtained 1042 instances of
agreement (98.3\%) and 18 cases of disagreement (1.69\%).

For comparison, the results obtained by comparing the second rater (whose input was not used for training) 
with the ten models obtained via ten-fold cross-validation described above we obtained the numbers of 
disagreements ranging from 11 (1.03\%) to 18 (1.69\%) (a mean of 15.18). This result is better than the agreement between the two raters. 

We thus have established the degree to which the two raters agree on
the decision, but not necessarily that either of the raters was correct.  To validate rater's opinions we,
therefore, administered a survey to a randomly selected set of authors. 
The survey provided respondents with a set of commits
with distinct author strings. All commits, however, were predicted
to have been done by the respondent and each respondent was asked to
indicate which of the commits were the ones made by them.  From a
randomly selected 400 developers sixty-nine emails bounced due to
the delivery problems. After 20 days we obtained 45 valid responses,
resulting in a response rate of 13\%. No respondents indicated that
commits predicted to be theirs were not submitted by them, for an
error rate of 0 out of 45. This allows us to obtain the bound on the magnitude of error. For example, if the algorithm has the 
error rate of 5\%, then we would have less than one in ten chances to observe 0 out of 45 observation to have errors\footnote{Assuming independence of observations and using binomial distribution.}. 

After establishing high accuracy of the training data we proceed to 
compare our approach to an approach that was implemented 
by professional commercial effort. 

\vspace{-2mm}
\subsection{Comparison with a commercial effort}
\label{bitergia}
Openstack is developed by a group of companies, resulting
in an individual and collective interest in auditing the development
contribution of each firm working on Openstack.  This task was
outsourced to Bitergia\footnote{https://bitergia.com/}, which is a
company dedicated to performing software analytics. We collected the
disambiguation data on OpenStack authors produced by Bitergia. The
data was in a form of a relational (mysql) database that had a tuple
with each commit sha1 and developer id and another table that
mapped developer id (internal to that database) to developer name
(as found in a commit). The Bitergia data had only 10344 unique author IDs that were mapped to
8,840 authors (internal database IDs). We first restricted
the set of commits in our dataset to the set of commits that were in
Bitergia database and selected the relevant subset of authors (10344
unique author IDs) from our data for comparison to ensure that 
we are doing the comparison on exactly the same set of authors.  
Bitergia algorithm misses 17,587 matches predicted by our algorithm and 
introduced six matches that our algorithm does not predict. 
In fact, it only detected 1504 matches of over 22K matches (under 7\%) 
predicted by our algorithm. Bitergia matching predicted 8,840 distinct 
authors or 41\% more than our algorithm which estimated 6,271 distinct authors. As shown in Table~\ref{tab:resultComp}, it has almost 50 times higher splitting error than manual classification, though it almost never 
lumps two distinct authors. We, therefore, conclude that 
the prediction done by the commercial effort was highly 
inaccurate.

\vspace{-2mm}
\subsection{Comparison with a research study}
Next, we compare our method to a recent research method\footnote{https://github.com/bvasiles/ght\_unmasking\_aliases} 
that was applied on data from 23,493 projects~\cite{msrData15} from GHTorrent to study social diversity
in software teams.  We refer to that method as ``Recent''. Method Recent
creates a record containing elements of each name and email address, 
forms candidate pools of addresses linked by matching name parts, then uses a heuristic to accept or reject each pool based on counts of different similarity ``clues''. The authors then iteratively adapted this automatic identity matching by manually examining the pools of matched emails and adjusting the heuristic.
To ensure that the heuristics in Recent were applied in a way consistent with their prior use, we asked the first author of the original paper~\cite{msrData15} to run it on our datasets and adjust it analogously to how he had adjusted for his own studies\footnote{As expected, the author got much better results than we could using his published code unmodified.}.
We first compare the results of Recent to our approach on the entire set of 16K OpenStack authors and then on a larger dataset described below.
As shown in Table~\ref{tab:resultComp}, Recent performs much better 
than the commercial effort. Splitting error is five times smaller than 
in the commercial effort, though it is approximately nine times higher than the manual matching (between the two raters). Lumping error is approximately three times higher than manual matching. Since ALFAA achieved errors that were even lower than manual matching, in fact suggesting valid corrections to manual matching as described above, we conclude that ALFAA exceeds the accuracy of Recent. In particular, it's  five times more accurate than Recent with respect to splitting, and three times with respect to lumping.

{\small
\begin{table}
\caption{Comparison of ALFAA against others}
\label{tab:resultComp}
\begin{adjustbox}{width=\linewidth}
\centering
\begin{tabular}{l l l l l l}
\toprule
\tabhead{Set} & \tabhead{Comparison} & \tabhead{Precision} & \tabhead{Recall} & \tabhead{Split} & \tabhead{Lump} \\
\midrule
Training & R1 vs R2 &  0.9861 & 0.9861 & 0.0139 & 0.0139  \\
Set & ALFAA vs ALFAA & 0.9990 & 0.9970 & 0.0030 & 0.001 \\
 & ALFAA vs R2 & 0.9936 & 0.9823 & 0.0177 & 0.0063 \\
\hline
Full & Bitergia vs ALFAA &  0.9991 & 0.4688 & 0.5312 & 0.0004 \\
OpenStack & Recent vs ALFAA & 0.9480 & 0.8891 & 0.1109 & 0.0487 \\
\bottomrule\\
\end{tabular}
\end{adjustbox}
\vspace{-8mm}
\end{table}
}
\newcommand{\squeezeup}{\vspace{-2.5mm}}
\subsection{Evaluation on a large set of identities}
To evaluate the feasibility of ALFAA on large scale we created 
a list of 1.8 million identities from commits to
repositories in Github, Gitlab and Bioconductor for
packages in 18 software ecosystems. The
repositories were obtained from libraries.io data~\cite{andrew_nesbitt} for
the Atom, Cargo, CocoaPods, CPAN, CRAN, Go, Hackage, Hex, Maven, NPM,
NuGet, Packagist, Pypi, and Rubygems ecosystems; extracted from
repository websites for Bioconductor\footnote{https://www.bioconductor.org},
LuaRocks\footnote{https://luarocks.org} and Stackage\footnote{https://www.stackage.org/lts-10.5}, and from Github searches for Eclipse plugins.

The application of Recent algorithm mapped the 1,809,495 author IDs to
1,411,531 entities as the algorithm was originally configured (1.28
aliases per entry), or
1,052,183 distinct entities after the heuristic was adjusted by its
author, identifying an average of 1.72 aliases per entry.
Upon applying our own algorithm to this dataset, we mapped the set to
988,905 --- identifying an average of 1.83 aliases per entity.
This indicates a similar ratio of aliases to entities to well-regarded
recent research approaches.
It is important to note that  we did not incorporate any additional training beyond the original set of manually marked pairs and we expect the accuracy to increase further with an expanded training dataset. 

Notably, to apply ALFAA for 1.8M IDs, we need $3.2\times 10^{12}$ string comparisons for each field (first name, last name, etc) and the same number of comparisons for each behavioral fingerprint. The full set of engineering decisions needed to accomplish the computation and prediction is beyond the scope of this paper, but the outline was as follows. 
To compare strings we used an allocation of 1.5 million core hours for Titan supercomputer 
at Oak Ridge Leadership Computing Facility (OLCF)~\footnote{https://www.olcf.ornl.gov/}. The entire calculation was done in just over 
two hours after optimizing the implementation in pbdR~\cite{pbdr} on 4096 16-core nodes. The approach can, therefore, scale to the entire set of over 23M author IDs in over 1B public commits. 
To compare behavioral fingerprints we exploited network properties (authors touch only a small number of all files) to reduce the number of comparisons 
by several orders of magnitude. Finally, it took us approximately two weeks to train 
Doc2Vec model on approximately 9M developer identities and 0.5B commits using Dell server with 800G RAM and 32 cores. 

\section{MEASURING IMPACT ON DEVELOPER COLLABORATION NETWORK}
\label{networkError}

\begin{figure}[htb]
\centering
\includegraphics[width = \linewidth]{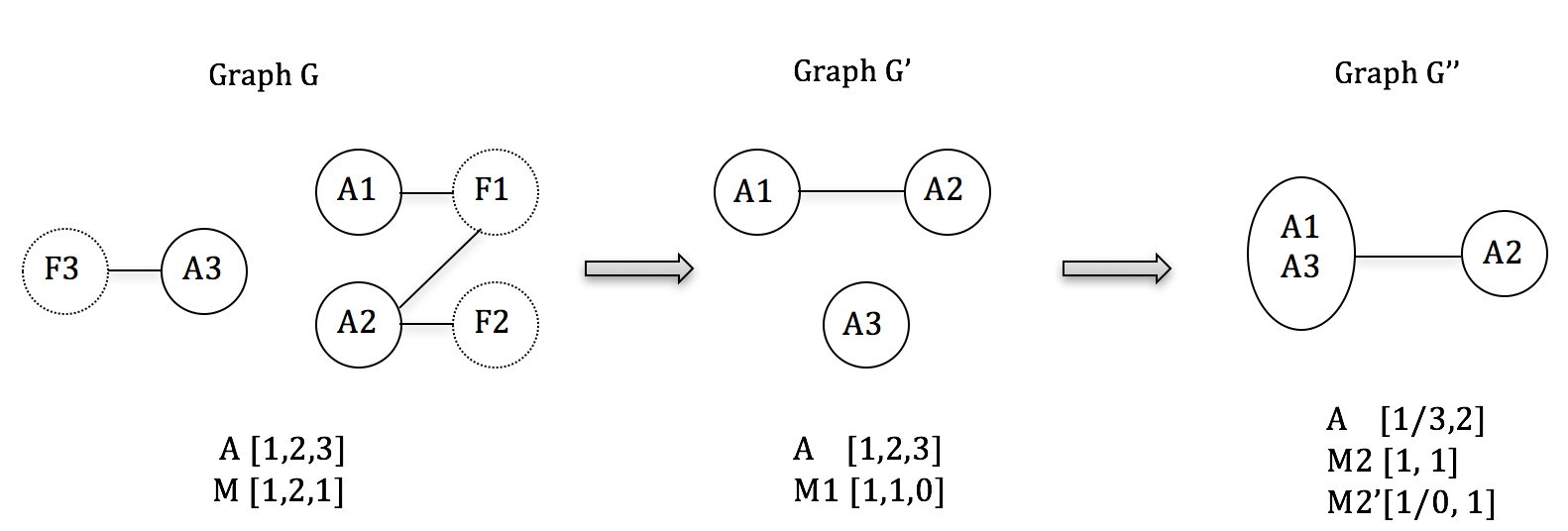}
\caption{Correcting Disaggregation Errors in a Developer Network} 
\label{devCol}
\end{figure}

In this section of our work, we discuss RQ5, the impact of identity
errors in a real world scenario of constructing a developer
collaboration network. More specifically, we measure the impact of
disaggregation (or split) errors by comparing the raw network to its
corrected version. To create the collaboration network (a common
network used in software engineering collaboration
tools~\cite{cataldo2006identification}),  we start from a bipartite
network of OpenStack with two types of nodes: nodes representing
each author ID and nodes representing each file, we refer to as G. The edges
connecting an author node and a file node represent the files modified by the author. 
This bipartite network is then collapsed to a regular author
collaboration network by creating links between authors that
modified at least one file in common. We then replace multiple links
between the authors with a single link and remove authors' self
links as well.   
The new network, which has 16,007 author nodes, depicts developer
collaboration, we refer to as G\textprime.  
We apply our disambiguation algorithm on G\textprime  and aggregate
author nodes that belong to the same developer and produce a
corrected network which we refer to as G\textprime\textprime. The
network and its transformations are illustrated in
Figure~\ref{devCol}.  

To evaluate the impact of correction from G\textprime  to
G\textprime\textprime, we follow prior work investigating the impact
of measurement error on social network
measures~\cite{wang2012measurement}. We look at four node-level
measurements of network error, i.e.   
degree centrality~\cite{FREEMAN1978215}, clustering
coefficient~\cite{cluscoeff}, network constraint~\cite{netcon} and
eigenvector centrality~\cite{EigenCen}. For each node-level measure
we compute a vector M. For example, in Figure 2, vector M has the
degree centrality of G, G\textprime \ and
G\textprime\textprime. Similar to the approach discussed
in~\cite{wang2012measurement} we compute two vectors M2 and
M2\textprime \ for graph G\textprime\textprime using each node level
measure and compute Spearman's rho between these two vectors. We obtain Spearman's rho for degree centrality to be $0.8619$, clustering coefficient to be $0.8685$, network constraint to be $0.8406$ and eigenvalue centrality to be $0.8690$.
The correlations below 0.95 for any of these measures are considered to indicate major 
disruptions to the social network~\cite{wang2012measurement}. In our case all of these measures 
are well below $0.95$. We can also look at the quantiles of these measures: for example one quarter of 
developers in the corrected network have 210 or fewer peers, but in
the uncorrected network that figure is 113 peers.  
The eigen-centrality has an even larger discrepancy: for one quarter of developers it is below 
0.024 for the corrected and 0.007 (or more than four times lower)
for the uncorrected network.   

\vspace{-2mm}
\section{Discussion}
\label{conclusion}

We have proposed a new approach (ALFAA) for correcting identity errors in software engineering context that is several times more 
accurate than a commercial effort and a recent research method
on OpenStack data. More importantly, the method does not rely on 
hand-crafted heuristics, but can, in contrast, we can increase its accuracy by simply adding more validated training data. In fact, it is designed to utilize the minimum amount of manual validation effort through active learning.


In order to answer RQ1, by examining a very large collection of commits we found that the identity errors were substantially different from the 
types of errors that are common in domains such as administrative records (drivers licenses, population census), publication
networks, or patent databases. While the data appears to have fewer phonetic spelling errors, it does contain similar typos. Additional errors involve template names or usage of names that imply desire for anonymity as well as missing data. Furthermore,the fraction of records with error appears to be much higher than in the other domains.

To answer RQ2 we summarize additional code commit information as behavioral fingerprints or vector embeddings of  the very high-dimensional space represented by files modified, the times of these modifications, and the word embeddings of the commit messages.  Such behavioral fingerprints 
provide information needed to disambiguate common instances of homonyms due to tool templates or desire for anonymity. 
As illustrated in Figure~\ref{workflow}, the homonym identification step involves high frequency fields and it also 
involves the cluster disaggregation stage. As noted in Section~\ref{relatedWork}, additional methods that detect high intensity of 
commits could also be used to identify additional potential homonyms. Once homonyms are identified, we would then go over each commit containing the homonym and search for developers that have fingerprints most closely matching the fingerprint of that commit.

To answer RQ3, we propose and implement an iterative approach
whereby the additional manual validation is conducted only on the
areas where the classifier has high uncertainty.
We found this
approach to rapidly lead to very high accuracy.
We established the feasibility to apply our approach to datasets 
that are on the order of the entire collection of author IDs in public repositories despite the computational intensity of the approach 
(need to calculate $O(n^2)$). 


To answer RQ4 we compared of our disambiguation approach with a commercial effort and with a recent research method. 
We found that our approach yield several times lower errors, 
suggesting that it does represent a real improvement over the 
state of practice.  Finally, to answer RQ5, we assessed 
the impact of measurement errors on the resulting networks. We found that use of 
uncorrected data would lead to major differences in resulting networks, thus raising questions about the validity of results for  
research that relies on such networks. 

We have produced a set of scripts and a model in an open source repository (the reference is omitted to 
comply with anonymous review requirements) that can be made more accurate simply with an addition of more training data.
We hope that the proposed method and associated tool will make it easier to conduct research and to build tools that rely on accurate identification of developer identities and, therefore, lead to future innovations built on developer networks.




\end{document}